# Experimental study on edge energetic electrons in EXL-50 spherical torus

Dong Guo*, Yuejiang Shi*, Wenjun Liu, Yunyang Song, Tiantian Sun, Bing Liu, Yingying Li, Xiaorang Tian, Guosong Zhang, Huasheng Xie, Y.K.Martin Peng, Minsheng Liu

Hebei Key Laboratory of Compact Fusion, Langfang 065001, China

ENN Science and Technology Development Co., Ltd., Langfang 065001, China

*Email: shiyuejiang@enn.cn   guodongd@enn.cn

**Abstract**

A significant number of confined energetic electrons have been observed outside the Last Closed Flux Surface (LCFS) of the solenoid-free, ECRH sustained plasmas in the EXL-50 spherical torus. Several diagnostics have been applied, for the first time, to investigate the key characters of energetic electrons. Experiments reveal the existence of high-temperature low density electrons, which can carry relatively a large amount of the stored energy. The boundary between the thermal plasma and the energetic electron fluid appears to be clearly separated and the distance between the two boundaries can reach tens of centimeters (around the size of the minor radius of the thermal plasma). This implies that the Grad-Shafranov equilibrium is not suitable to describe the equilibrium of the EXL-50 plasma and a multi-fluid model is required. Particle dynamics simulations of full orbits show that energetic electrons can be well confined outside the LCFS. This is consistent with the experimental observations.

## I. Introduction

The toroidal plasma current is essential to keep the plasma in equilibrium and to form nested magnetic surfaces for confinement in tokamaks. Usually, this magnetic equilibrium can be described by a single-fluid model and reconstructed by using the Grad-Shafranov formalism with the assumption that the dominant plasma current is contained inside the LCFS[1]. However, this assumption is no longer valid when a significant number of confined energetic electrons exist outside LCFS and carry a considerable plasma current. In order to describe the equilibrium of such a type of plasmas, several models have been developed[2-4]. A three-fluid (two electron fluids and one ion fluid) axisymmetric equilibrium model with toroidal and poloidal flows was developed and firstly applied to TST-2 spherical tokamak [2]. It was found that the toroidal current density and pressure are dominated by the low-density high-energy electron fluid and the radial force balance for each fluid species is quite different. A four-fluid axisymmetric plasma equilibrium model was proposed for further consideration of the relativistic effects of energetic electrons[3]. There are also many experimental studies on energetic electrons. ECRH power modulation Experiments on TST-2 showed that large number of energetic electrons in SOL can cause the floating potential to change from negative to positive[5]. A co-directional toroidal flow of roughly 4 keV energetic electrons was detected on QUEST[6]. Such energetic electrons can carry a significant current of about 2–3 kA/m². However, few studies have been reported on the experimental validation of the multi-fluid equilibrium.

In order to study the characteristics of the energetic electrons, several dedicated experiments were performed in EXL-50. Firstly, two different material tips, i.e., stainless steel and tungsten, were mounted on a Langmuir probe to monitor the electron emission effect under the impact of energetic electrons. In addition, a boron powder injection experiment was conducted to study the relation between the boundaries of the main plasma and energetic electrons. Lastly, the edge threshold heat flux causing probe melting was determined from thermodynamic simulations, indicating the energy range of energetic electrons for different density assumptions. Full orbit simulations based on the reconstructed magnetic equilibrium from a multi-fluid model was also performed, showing that energetic electrons can be confined outside the LCFS.

In Sec. II, a brief introduction of EXL-50 and experimental setup is introduced. Detail experimental results and comparisons with simulations are presented in Sec III. A discussion and a summary are presented in Sec IV.

**II. Experimental Setup**

EXL-50 is a medium-size Spherical Torus (ST) [7], built in 2019 at the ENN Energy Research Institute. One of the key EXL-50 experimental goals aims to test the efficiency of the electron cyclotron resonance heating and current drive (ECRH&CD) in the absence of a central solenoid[8]. Its TF and PF coils are conventional copper conductors which allow for a plasma discharge duration of ~10 seconds. The vacuum vessel (VV) is made of 316L stainless steel with the major radius R =1.65m. At present, EXL-50 is equipped with two 400kW ECRH heating systems at a frequency of 28GHz and more than ten plasma diagnostics.

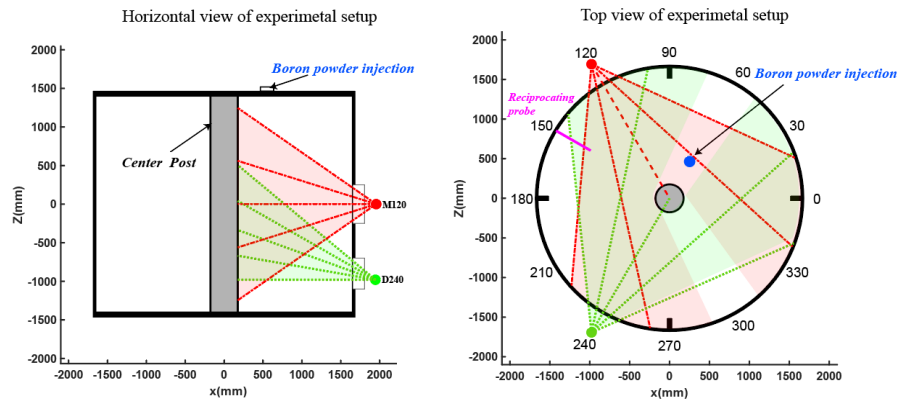

Figure 1. Different views of EXL-50 VV and positions of the key diagnostics used in this paper. Left is the poloidal projection of the cameras and their field of views. Right is the bird-view of the diagnostics. The reciprocating probe is located on the mid-plane at 150 degrees. The boron powder injection port is at the top of the VV at 60 degrees.

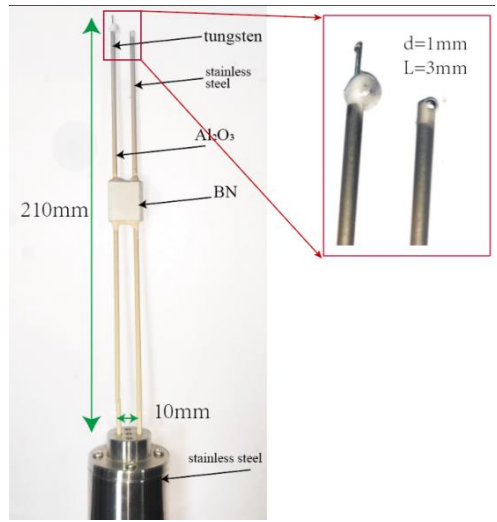

Figure 2. Photograph of the Langmuir probe with the different material tips. The zoomed-in area shows the burned probe tip after the experiment (left is tungsten and right is stainless steel).

Generally, the information of the energetic electrons inside the LCFS can be obtained from the hard x-ray (HXR) diagnostic[9], while the Langmuir probe can give some quantitative measurements of energetic electrons outside the LCFS taking into account the secondary electron emission and electron reflection from the probe [10]. The key diagnostics used for this study are a mid-plane Langmuir probe and 2 high speed visible light cameras. The two monochrome cameras, Phantom V1212 model, are equipped with 1280×800 pixels viewing the plasma with different views. The spectral response is from 350nm to 1050nm with the peak response at 700nm. As shown in figure 1, an M120 camera installed on the mid-plane at 120 degrees provides a full view of plasma geometry. A D240 camera installed on the lower flange at 240 degrees provides an upper-side-view. The reciprocating system is placed on the mid-plane at 150 degrees, which can be viewed by the D240 camera if the probe penetrates sufficiently deep into the plasma. A set of boron powder injection apparatus is installed at the top of VV at 60 degrees. 70um diameter pure boron powders were introduced gravitationally into plasma discharges at rates of 8-10mg/s. In this study, full resolution is used at a rate of 300 frames per second. The Langmuir probe (shown in figure 2) is mounted on the reciprocating probe system with the maximum speed is 0.47m/s using a server motor and ~1.5m/s using a cylinder drive. In order to monitor the effect of the secondary electron emission at different materials, two adjacent probes spaced 10 mm from each other, are made of stainless steel and tungsten respectively. In order to penetrate deeper into the plasma without affecting the plasma discharge, a small probe head with 1 mm diameter and 3mm height out of the $Al_2O_3$ insulation tube was employed. The total length of the thin tube is about 210 mm. A Boron nitride support is placed in the middle of the long $Al_2O_3$ tube to ensure that the two tubes are parallel.

## III. Experimental results
### A. Observation of energetic electrons outside the main thermal plasma region

Figure 3 shows the time evolution of the typical plasma parameters of a hydrogen plasma discharge (shot 9695) with 120kW ECRH power being applied from t=0~4.5s. The absolute power of the ECRH is obtained at the matching optical unit close to the gyrotron by the calibration of the water load daily before the operation. The probe is triggered at t=1.8s and penetrates into the plasma from R-$R_{LCFS}$=10cm with an average speed of 0.23m/s. At 3.3s, the probe penetrates to the deepest position at R-$R_{LCFS}$=-20cm. The probe stays still about 0.1s in the plasma from 3.3~3.4s, as marked by the blue shadow area. The probe is then returned to the original position at the same speed. During the whole process, the probe penetrates in and out without significant impacts on the plasma current $I_P$, which changes slowly from 90kA to 130kA with the variation of the poloidal field (PF) coil current. As shown in figure 3(b), the line-integrated plasma density shows little changes during the phase of probe penetration before 3.3s. After that, it increases exponentially from 8 to $12\times10^{17}$/m². The hard X ray increases as the probe enters the plasma; the counts of the hard-x ray detector are higher with larger probes (tips and support structure) being exposed to the plasma. This is due to thick-target bremsstrahlung emissions caused by the energetic electron bombardment of the probe. When energetic electrons interact with the probe, it is generally necessary to consider the effect of electron emissions from the probe on the probe measurements. The measured current is actually the difference between the current entering the surface of the probe and the electron current emitted from the probe. The floating potential $V_f$ can be finally expressed by the following expression

$$V_f - V_{\text{plasma}} = \frac{T_e}{2e} \ln\left[(1-\xi)^{-2} \frac{2\pi m_e}{m_i} \cdot \frac{T_e + T_i}{T_e}\right] \quad (1)$$

$V_{plasma}$ is the plasma potential, $m_i$, $m_e$, $T_i$ and $T_e$ are the mass and temperature of plasma ions and electrons, respectively. From equation 1, it can be seen that the measured value $V_f$ is more positive when the electron emission coefficient $\xi$ is nonzero. The total emission coefficient $\xi$ is defined as $\xi = \delta + \eta$, where $\delta$ is due to the secondary electrons while $\eta$ represents the reflected or back-scattered electrons. Note that the total secondary electron emission coefficient of tungsten is larger than that of stainless steel[11]. This may explain the experimentally observed differences in the probe measurements for different materials.

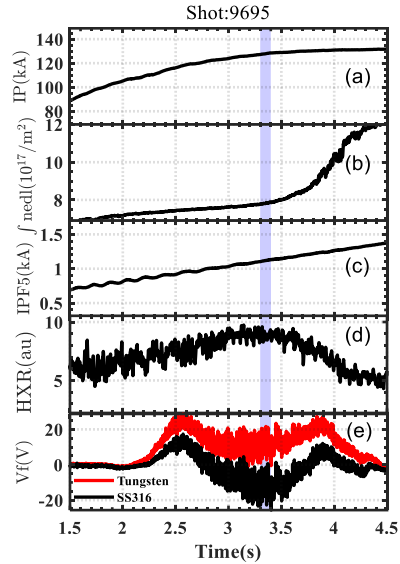

Figure 3. Time evolution of plasma parameters: (a) plasma current IP (b) line integrated density (c) impurity concentration (d) hard x-ray intensity and (e) floating potential.

The process of probe in- and out-penetration can be monitored by the D240 camera which provide the upper-side-view of the plasma and the probe. The ultra-bright spot on the image mainly comes from the continuous visible light emitted on the melting head of the probe. At 2.85s, the probe comes into the camera view with a very small burned area, as seen on the image. The probe stops moving at 3.3s and the burned area becomes larger. The noises appearing on the CCD image results from the impact of hard x-ray on the camera CCD detector. The number of white noises peaks at t=3.3s, which indicates that the magnitude of hard x-ray at this time is much stronger than other times. This is consistent with the hard x-ray measurement in fig.3. This indicates that there exist a large number of energetic electrons outside the main plasma boundary, causing the probe to be burned and melted during its reciprocating process.

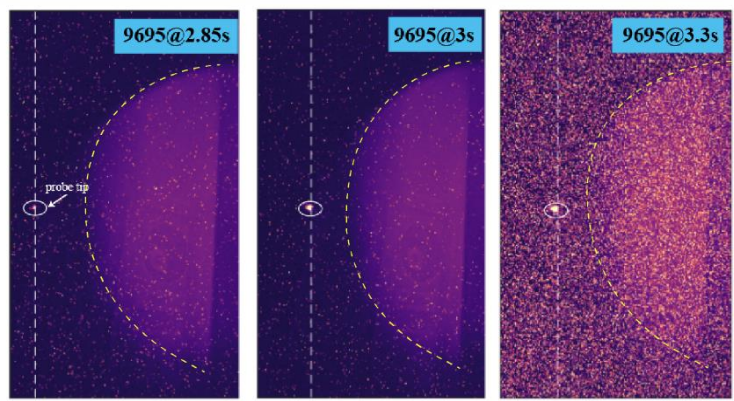

Figure 4. Images of probe penetration into the plasma at different times from the D240 camera. The white dotted line indicates the position of probe penetrated into the plasma and the yellow line means the main plasma boundary.

Another experiment was conducted to verify the relation between the main plasma boundary and that of the energetic electrons. Figure 5 shows the images during the boron powder injection process, observed by the M120 camera at different times. The powder drops into plasma as dust and it is a complex process including interaction, ionization and charge exchange with the background plasma[12]. A clear bright light can be observed near the injection port of boron powders. The plasma shape is gradually compressed by adjusting the PF coil current to distinguish the boundary of the main plasma and energetic electrons. The images indicate that with a certain injection rate, the boron powders are completely ionized before reaching the maximum depth, as indicated by the green dashed line. The white dotted line shows the initial position of the powders being ionized. The injection depth of boron powders for this discharge is about 0.5m. Thus, we can estimate the boron atoms $\Gamma$ within the injection depth[13].

$$\Gamma = d/\upsilon \cdot r/M_B \cdot N_A \quad (2)$$

In this equation, $d$ is the injection depth 0.5m, $\upsilon$ is the mean velocity of boron atoms 6.87m/s, $r$ means the boron (mass) injection rate. $M_B$ is the molecular mass of boron and $N_A$ being Avogadro constant. Under such conditions, the number of boron atoms is about $4 \times 10^{19}$ and its density is larger than $4 \times 10^{19}$ /m$^3$ with the volume less than 1 m$^3$ within the injection depth. This would be much larger than the value of main plasma density $8 \times 10^{17}$/m$^3$. Low temperature and low density plasma outside LCFS may not be able to ionize such a large number of boron atoms. Therefore, we can infer that the ionization of the boron atoms may be due to the numerous collisions with energetic electrons, which needs to be further investigated.

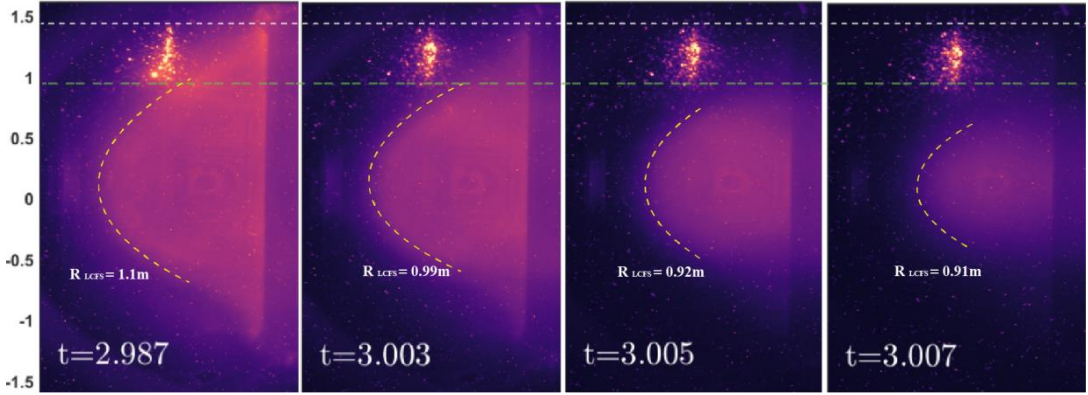

Figure 5. The injection process of boron powders captured by the M120 camera during the discharge (shot #12094). The white dashed line indicates the start position of boron powder drop due to the gravity and the green dashed line indicates the stop position.

**B. Reconstruction of the plasma equilibrium**

In tokamaks, the LCFS position is routinely calculated using a Grad–Shafranov MHD equilibrium solver, such as EFIT, if adequate plasma information is supplied. However, it fails when a significant plasma current exists outside LCFS. Recently, a multi-fluid equilibrium reconstruction model[3] has been developed for this special plasma scenario. The calculated results are in good agreement with the experimental measurements, such as plasma density $n_e$, temperature $T_i$, current $I_p$, magnetic flux $\Phi$ and so on. In addition, the high speed visible light camera can give us a clear boundary of the main plasma. Many previous studies indicate that the optical boundary can reflect the actual LCFS, and the error is generally less than 10 cm[14, 15]. Such a method is also adopted to determine the LCFS in EXL-50[16].

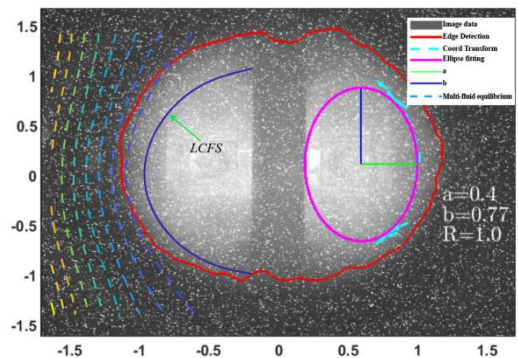

Figure 6. Optical reconstruction and multi-fluid simulation results for shot #9695@3s. The dashed lines on the left side of the center post shows the flux surface and the right shows the fitting result from the optical reconstruction. The red circle shows the plasma boundary brightness threshold contour and its coordinate transformed result is shown in cyan line. The ellipse fitting result of transformed data is shown in magenta line.

Figure 6 shows the plasma geometry for shot #9695@3s. The boundary threshold of the plasma-emitting region is obtained by fitting the brightness histogram of the image, which is shown as the red line. Assuming that the plasma is a spherical luminous body, the CCD image is its projection in the two-dimensional plane, a coordinate transformation is required to reflect the true coordinates of the plasma[17]. The coordinate-transformed curve is shown in cyan line, which can tell us the rough estimation of the plasma major radius. Further, given the boundary of the central post, we can obtain the long and short axes of the plasma by eclipse fitting. The optical reconstruction of the plasma geometry is shown on the right of the CCD image and on the left is the simulation result obtained by the multi-fluid code. The blue solid line is the LCFS at the major radius of 0.96m. The dotted line out of the LCFS represents the open flux surface. Figure 7 shows the plasma density and temperature profiles obtained from the multi-fluid simulation. The result shows that energetic electrons exist both inside and outside the LCFS. The temperature of energetic electrons is almost three orders of magnitude larger than that of the main plasma, while its density is one magnitude smaller than that of the main plasma. The boundary of thermal plasma (LCFS) in the middle plane is around R=0.96m while the boundary of energetic electrons is at about R=1.2m.

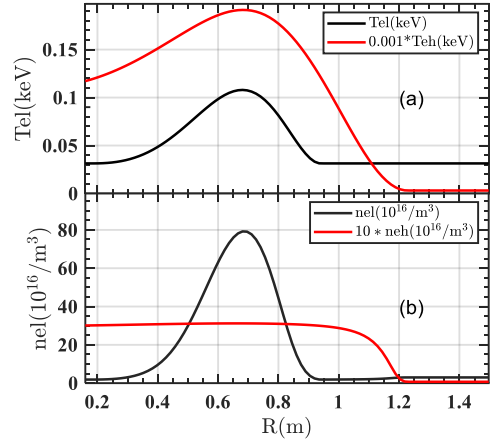

Figure 7. Plasma density and temperature profiles calculated by the multi-fluid code. (a) Temperature profiles of thermal electrons and energetic electrons. (b) Density profiles of thermal electrons and energetic electrons.

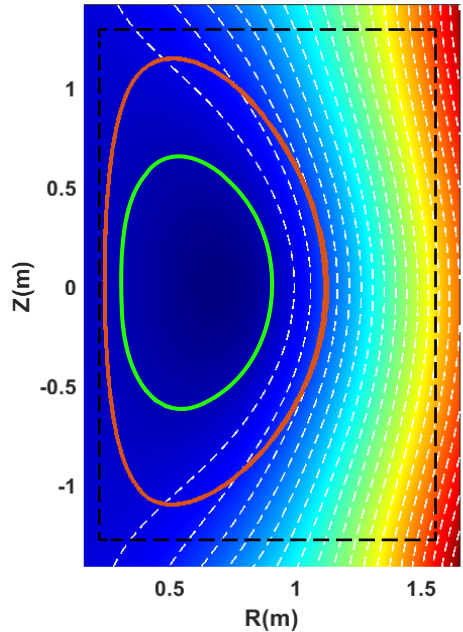

Figure 8. The simulation of the confined-orbit region of the energetic electrons using the equilibrium magnetic field calculated for a plasma discharge (#9695@3s). The white dashed lines represent the flux surfaces and the green line indicates the LCFS. The black dashed line indicates position of the limiter wall of EXL-50. The brown solid line shows the passing orbit of 2MeV energetic electrons.

    A particle orbit simulation can provide a "first-principle" physical picture of the energetic electrons produced in the EXL-50 experiment. As shown in figure 8, the contour plot shows the equilibrium magnetic flux with the white dashed lines representing the open flux surfaces and the green solid line indicating the LCFS. The black dashed line shows the position of the limiter boundary that constrains the plasma. The brown line shows the passing orbit of 2MeV energetic electrons confined outside the LCFS. The orbit simulation, by scanning the initial position of the particles, shows

that the maximum radial position of the particles that can be confined can reach R=1.13m, and the energy of the particles that can be confined at this position can reach up to 2 MeV.

## C. Experimental evaluation of energetic electrons temperature

Despite some uncertainties involved in the probe data analysis, Langmuir probe is one of the primary means of diagnosing the plasma boundary. The plasma facing surface of the probe may be subject to large heat fluxes. The parallel heat flux $q_\parallel$ to the target surface is given by [18, 19]

$$q_\parallel = q_k + q_p = (\gamma T_t + \varepsilon_{pot})j_s/e \quad (3)$$

Where $q_k = \gamma n_e C_s T_t$ and $q_p = n_e C_s \varepsilon_{pot}$ are the kinetic and potential heat fluxes, respectively, $C_s$ is the ion sound speed, $j_s$ is the ion saturation current, and $\varepsilon_{pot}$ is the potential energy deposited on the surface per incident ion, which can be negligible compared with the kinetic power when $\varepsilon_{pot}/\gamma T_t \ll 1$, $\gamma$ is the sheath transmission coefficient, which relates $q_\parallel$ to the electron temperature $T_e$ and particle flux as $\Gamma_s = n_e C_s$ at the sheath edge.

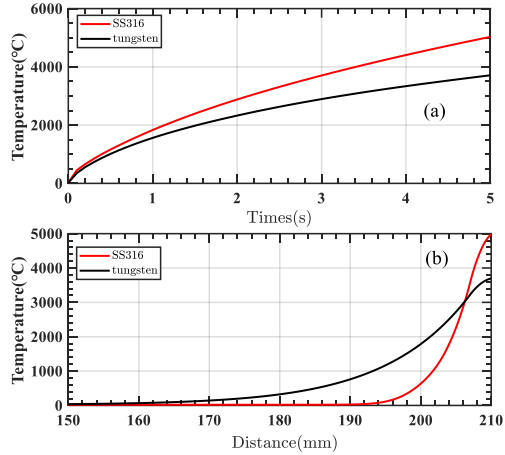

Figure 9. Calculations for probe tip (3mm length, 1mm diameter) exposed to 4.2 MW/m² heat flux for 5s. (a) Time evolution of the maximum temperature at the probe tip, and (b) the temperature distribution along the probe. The top of the probe tip is at 210 mm.

Experimental results show that a single discharge can cause the melting of the tungsten probe tip when the reciprocating probe moves from R-R$_{LCFS}$=0~-20cm. The melting point of tungsten is 3410 °C. Assuming no thermal convection between the plasma and the probe, the temperature rise induced by the heat load on the probe surface can be calculated with heat conduction. Figure 9 (a) shows the simulation result of the time evolution of the maximum temperature at the probe tip exposed to 4.2 MW/m² heat fluxes. As can be seen, the temperature of tungsten probe rises to the melting point within 5 seconds. The temperature rise of stainless steel can reach as high as 5000°C, which is much higher than its melting point at 1850°C. The picture of the exposed probe tips in figure 2 clearly shows stronger melting on the stainless steel probe tip than on tungsten probe tips. Figure 9(b) shows the temperature distribution along the probe. The calculation results show that the temperature decays exponentially from the top of the probe tip, down to room temperature 5 centimeters away. Hence, the melting of the probe tips, as shown in fig.2, could happen when the heat load on the probe tips is higher than 4.2 MW/m$^2$.

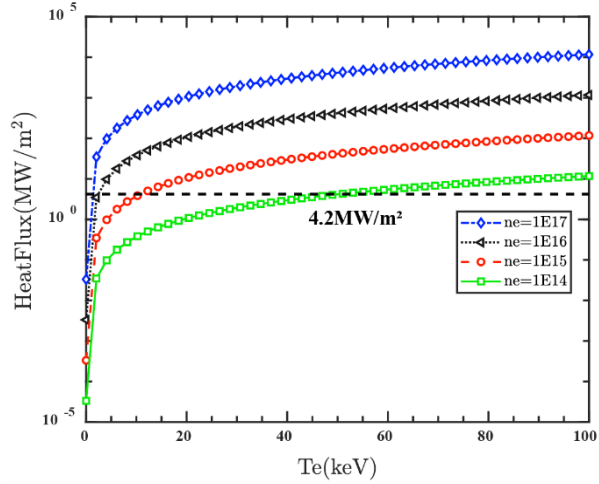

Figure 10. Heat flux on the tungsten probe versus temperature of energetic electrons for different densities.

The relation between the heat fluxes to the probe tip, $q_t$ and $q_\parallel$, is expressed as following

$$q_t = q_\parallel \cdot sin\theta = q_\parallel \cdot B_p/B_T \quad (4),$$

where $\theta$ represents the angle between the poloidal magnetic field $B_p$ and the toroidal magnetic field $B_T$. The multi fluid simulation result shows that in the radial region where probe moves from R=1.525m to R=1.325m, the ratio of $B_p$ to $B_T$ is about 0.24. The density and temperature of the thermal plasma at LCFS in EXL-50 are less than $10^{17}$m$^{-3}$ and 30eV, respectively, which should be lower outside the LCFS. With the assumption that $T_e=T_i$, we can estimate the heat flux of the main plasma as $q_\parallel = \gamma n_e C_s T_t \cong 0.155 \cdot ne[10^{19}/m^2] \cdot Te^{1.5}[eV]$ in MW/m$^2$. Thus, the heat flux of the main plasma to the probe tip $q_t$=0.06MW/m$^2$, which is much smaller than the threshold value for the melting of tungsten or stainless steel. Since there are no measurements of temperature and density of energetic electrons outside the LCFS, we can reasonably estimate the parameters of energetic electrons based on the heat flux from probe melting. Figure 10 shows the heat flux threshold to cause the melting of the tungsten probe tip as a function of temperature for different densities. Based on these calculations, the density of energetic electrons appears to be within the range of $10^{14}$~$10^{16}$/m$^3$ for the temperature of energetic electrons in the range of 2~45keV.

## IV. Summary and discussion

Experiments have been performed to verify the existence of confined energetic electrons outside the LCFS in the EXL-50 spherical tokamak. The measurements from Langmuir probes showed that the floating potential can be more positive due to the secondary electron emission, suggesting the presence of energetic electrons. The probe tips with different materials, i.e, tungsten and stainless steel, were subject to different degrees of melting. Simulation results showed that the probe tips are subject to the minimum heat load of 4.2 MW/m$^2$, which is caused by the energetic electrons with energy ranging from 2keV~10keV and density from $10^{15}$~$10^{16}$/m$^3$. In addition, a boron powder injection experiment was conducted to study the relation between the plasma boundary and energetic electrons. Further, the confined-orbit of the energetic electrons was simulated using the multi-fluid equilibrium magnetic field, showing that the energetic electrons can be confined outside the boundary of the main plasma.

## V. Acknowledgement

We thank Dr. Houyang Guo for his careful revise and helpful discussion for the experiment data analysis. We thank EXL-50 team for their nice operation of experiment.